\documentclass[preprint2]{aastex}%
\usepackage{url}

\usepackage{amsmath}
\usepackage{color}

\RequirePackage{color}

\begin{document}

\title{Modeling the Spatial Distribution of Neutron Stars in the Galaxy}

\shorttitle{Physical Characteristics of Accretion Induced Collapse}
\shortauthors{Taani et al.}

\author{Ali Taani, Luca Naso, Yingchun Wei, Chengmin Zhang and Yongheng Zhao}
\affil{National Astronomical Observatories, Chinese Academy of
Sciences, Beijing 100012, China} \email{alitaani@bao.ac.cn}


%

\begin{abstract}
In this paper we investigate the space and velocity distributions of
old neutron stars (aged $10^{9}$ to $10^{10}$~yr) in our Galaxy.
Galactic old Neutron Stars (NSs) population fills a torus-like area
extending to a few tens kiloparsecs above the galactic plane. The
initial velocity distribution of NSs is not well known, in this work
we adopt a three component initial distribution, as given by the
contribution of kick velocities, circular velocities and Maxwellian
velocities. For the spatial initial distribution we use a $\Gamma$
function. We then use Monte Carlo simulations to follow the
evolution of the NSs under the influence of the Paczy\'{n}ski
Galactic gravitational potential. Our calculations show that NS
orbits have a very large Galactic radial expansion and that their
radial distribution peak is quite close to their progenitors' one.
We also study the NS vertical distribution and find that it can well
be described by a double exponential low. Finally, we investigate
the correlation of the vertical and radial distribution and study
the radial dependence of scale-heights.
\end{abstract}


\section{Introduction}\label{sec:intro}
It is commonly accepted that Neutron Stars (NSs) are born when
massive OB-stars exhaust their nuclear fuel and end their lives in
core-collapse supernova explosions, near the Galactic disk (see e.g.
\cite{BH91}) and that they have then moved away from the Galactic
plane with average kick velocities around $200 - 500$~km/s (e.g.
\cite{cor98}, \cite{lyn04}, \cite{hob05}). About $10^{9}$ NSs are
thought to populate our Galaxy, but only ~$2\times10^{3}$ are
directly observed as radio pulsars or as accretion-powered X–ray
binaries \citep{sar10}, as a consequence, little is known about
their statistical properties. The estimation of pulsar velocities
relies on direct distance measurements, which can be obtained by
dispersion measures together with a Galactic electron density model
(e.g. \cite{Tay93}, \cite{CL2002}). The mechanisms for producing
high velocities are still unknown \citep{lor08}.

Numerical simulations are valuable tools for understanding the
spatial and velocity distribution of NSs in the Galaxy, and they
have been used by several authors (see e.g. \cite{cal81},
\cite{Ofe09}, \cite{kat11}). In particular, \cite{pac90} (hereafter
P90) simulated the motion of NSs in connection with the galactic
origin of gamma ray bursts and calculated the NS space density
distribution. In the same work Paczy\'{n}ski also suggested a
simplified expression for the gravitational potential which is still
often applied in the simulation of NS distribution. For example
\cite{wei10a} used this potential. They considered the old NSs, i.e.
NSs whose age is between $10^{9}$ and $10^{10}$~yr. They adopted a
Galactic distribution with one-component initial random velocity
models. The aim of the present work is to improve the model
developed in \cite{wei10a} for studying the distribution of old NSs
as a function of the initial position distribution, of the initial
velocity distribution and of the galactic gravitational potential.
Based on P90 gravitational potential we consider the evolution of a
two-component Maxwellian initial random velocity distribution (we
adopt the velocity distributions of NSs at birth from
\cite{ARal2002} and \cite{fau06}). We perform integration of NS
velocities using Monte Carlo integration techniques with different
conditions developed for this purpose. We also aim at obtaining the
NS trajectories under variety of assumptions.

The paper is structured as follows: in Sec.~\ref{sec:model} we
describe the ingredients of the simulation, i.e. the NS initial
position and velocity distribution and the Galactic gravitational
potential. We present the results of the simulation in
Sec.~\ref{sec:orb} and~\ref{sec:rad}. In Sec.~\ref{sec:orb} we
investigate the NSs orbits; while in Sec.~\ref{sec:rad} we
investigate their vertical and radial distributions. We fit the two
exponential decay model, for each spacing segment of R to derive the
high scale heights. We fit also the R distributions at different
scale heights. We discuss our results and their possible
implications in Sec.~\ref{sec:rest}.

\section{Simulation ingredients}\label{sec:model}
In this Section we present the ingredients of the model: the NS
birth rate and Monte Carlo simulation; the gravitational potential;
the position distributions of NSs; and the initial velocities and
equations of motion.

\subsection{NSs birth rate and Monte Carlo simulation}\label{sec:rate}
Theoretically, estimating the birthrate of a population of sources
is simple. However, for the NS population, precise estimates of both
the number and lifetime of the sources are hard to obtain, because
they may have been heavily biased by a number of observational
selection effects. The birthrate of NSs ($\eta$) within the whole
Galactic disk is roughly $0.9 \sim 2$ per century \citep{lyn07,
lor08}. If one assumes a life time $\tau \sim$ $10^{9}$ -
$10^{10}$~yr, then an estimate of the total number of NSs is:

\begin{align}
N = \tau\ast\eta \simeq  10^{7-8}
\end{align}

The problem is best tackled using a Monte Carlo simulations of NS
positions, orbits and velocities, taking into account the
birthplaces and the initial kick velocities. We study the resulting
phase spatial distributions concerning the Galactic potential and
the distribution of progenitors and birth velocities, focusing on
the numerical properties of the NS populations in the disk and in
the solar neighborhood. NS orbits are obtained by solving the
equations of motion in the P90's gravitational potential.

\subsection{Galactic gravitational potential}
It is known that the Galactic gravitational potential causes
oscillation of objects along the direction perpendicular to the
Galactic plane (see e.g. \cite{lyn82}, and references therein). To
track the evolution and motion of NSs population, the gravitational
potential P90 is taken to be a homogeneous function of the density,
and ignore the interstellar friction. This is a reliable
approximation for our axisymmetric model, because the steady state
distribution of old NSs depends only weakly on the non-homogeneous
part of the galactic potential \citep{fre92}. Asymmetry in the
kinematics, which is likely due to the finite lifetime of the stars
and Galactic potential structure, is a relatively small effect
\citep{Per09}. However, using P90 may not be a good approximation
when studying non-axisymmetric models, because rotating
non-axisymmetric components (like bar or spirals) can introduce
resonances (see e.g. \cite{Pat96}, \cite{Pat02}).

Our evolution calculations are presented to simulate more realistic
old NS distribution under the two-component Maxwellian initial
random velocity.  We model the gravitational potential of the Galaxy
following P90:
\begin{equation}
\label{eq:(1)}
  \Phi = \Phi_{\rm sph}+\Phi \rm _{disk}+\Phi \rm _{halo}\;,
\end{equation}
where $\rm \Phi_{sph}$, $\rm \Phi_{halo}$ and $\rm \Phi_{disk}$ are
the spheroid, halo and disk components, respectively.

For the spheroid and disk components one has:
\begin{align}
\label{eq:(2)}
  \Phi \rm _{i}(R,z)= -
\frac{GM_{i}}{\sqrt{R^{2}+[a_{i}+(z^{2}+b_{i}^{2})^{1/2}]^{2}}}
\end{align}
where $\rm R=\sqrt{x^{2}+y^{2}}$ is the distance from the Galactic
rotation axis and $\rm z$ is the distance from the Galactic disk
plane. The subscript ``i" represents ``sph" and ``disk". The values
for the parameters are taken from P90 and for the spheroid component
they are: a$\rm _{sph}=0.0$~kpc, b$\rm _{sph}=0.28$~kpc and $M_{\rm
sph} = 1.12 \times 10^{10}$~M$\rm _{\odot}$; while for the disk
component: a$\rm _{disk}=3.7$~kpc, b$\rm _{disk}=0.20$~kpc and M$\rm
_{disk}=8.01\times10^{10}$~M$\rm _{\odot}$.

The halo component of the Galactic gravitational potential is:
\begin{align}
 \label{eq:(3)}
\nonumber
 \Phi\rm _{halo}=\frac{GM\rm_{halo}}{r_{c}}
  \left[\frac{1}{2}\ln\left(1+\frac{R^{2}+z^{2}}{r_{c}^{2}}\right) + \right.\\
\left.\rm \frac{\rm r_{c}}{\rm \sqrt{\rm R^{2}+z^{2}}}\arctan
\left(\rm \frac{\rm \rm \sqrt{\rm R^{2}+z^{2}}}{\rm
r_{c}}\right)\right]
\end{align}

where $\rm r_{\rm{c}}=6.0$~kpc and $\rm M_{\rm{halo}} = 5.0 \times
10^{10}$~$\rm M_{\odot}$.

\subsection{NS initial position distribution}
It is generally accepted that the galactic z-distribution of massive
objects is approximately exponential \citep{bin98, Mdz04}. This kind
of the distribution can be theoretically explained by considering
the dynamic equilibrium within the Galaxy. The initial $\rm z$
probability density function of NSs in the Galaxy has been proposed
by \cite{gott70} and adopted by many authors since then (e.g.
\cite{gon02}):
\begin{equation}
\label{eq:(4)}
  \rm p_{z}(z)=~\frac{1}{2 h_{z}} \exp\left[\frac{-|z|}{h_{z}} \right]
\end{equation}
where h$\rm_z=0.07$~kpc is the scale height and:
\begin{equation}
 \rm \int_{0}^{\infty}\frac{1}{h_{z}}  \exp\left[\frac{-|z|}{h_{z}}\right] dz = 1
\end{equation}

For the initial radial probability density function of the NSs we
adopt the same expression as \cite{ARal2002}. As in P90, it follows
a gamma function $\Gamma(2,4.5)$, but has a radial outer boundary at
$15$~kpc rather than at $20$~kpc\footnote{\textbf{Because of the
rapid decrease of the Gamma function with $R$ we do not expect this
modification to have a large impact on the results.}}. This is
motivated by the radial distribution of NS progenitors, i.e.
population I massive stars. Although the Galaxy is believed to have
a stellar disk $0\sim15$~kpc and a gaseous disk $15\sim25$~kpc, NS
progenitors hardly form in the gaseous disk, due to the considerable
decrease of the gas density \citep{Jon04}.

The initial radial probability density function that we use is the
following:
\begin{align}
\label{eq:(5)}
  \rm p _{R}(R) = a_{r} \frac{R}{R_{\rm exp}^{2}} \exp \left[ -R/R_{\rm
exp} \right]\;,
\end{align}
where
\begin{align}
 \rm a_r = & \left[ 1 - \exp^{-\frac{R_{\rm max}}{R_{\rm exp}}}~\left( 1 +
\frac{ R_{\rm max}}{R_{\rm exp}} \right) \right]^{-1}\;.
\end{align}
We use $R_{\rm exp}=4.5$~kpc, which gives $a_r \simeq 1.183$. The
probability distribution is normalized to $1$ within the considered
radial domain, i.e. from $0$ to $15$~kpc.

\subsection{NS initial velocity distribution and equations of motion}

The NS initial velocity is \textbf{calculated as} the vector
addition of three different velocities: (1) a Maxwellian
distribution, (2) a constant kick, and (3) the circular rotation
velocities at the birthplace.

Maxwellian distributions are usually used to represent the observed
distribution of pulsar velocities. In this work we choose a
two-component Maxwellian distribution. One component includes $40\%$
of all NSs and has a velocity dispersion $\sigma_{v} \sim 90$~km/s.
The other one includes the remaining $60\%$ and has $\sigma_{v} \sim
500$~km/s, as proposed by \cite{ARal2002}.

While for the kick velocity we adopt the conventional value of about
400 km/s for every single object \citep{han97, hob05}.

The initial circular rotation velocity of the NS is determined by
\begin{equation}
\label{eq:(6)}
  \rm v_{circ}=\left(R\frac{d\Phi}{dR}\right)^{1/2},
\end{equation}
where $\Phi$ is the P90's gravitational potential in
Eq.~(\ref{eq:(2)}).

The differential equations that describe the NS  motion in the
Galaxy can be expressed in the compact vector form as
\begin{equation}
\label{eq:(7)}
  \rm \ddot{\vec{r}}=-\vec{\nabla}\Phi\left(\rm \sqrt{x^{2}+y^{2}},z\right),
\end{equation}
where r\rm $=\rm \sqrt{x^{2}+y^{2}+z^{2}}$ is the spherical distance
from the galactic center. NS orbits are numerically
integrated with the fourth-order Runge-Kutta method.\\

The NS total energy integral is used to control the accuracy of the
integrations and in our simulations the total energy change is less
than 1 part in $10^{6}$. The accuracy changes for different orbits,
and generally simpler orbits are more accurate.

\section{NS orbits in the Galaxy}\label{sec:orb}
The Poincar\'{e} section technique is a way of presenting a
trajectory in $(n)$-dimensional phase space in an
$(n-1)$-dimensional space. By picking one phase element constant and
plotting the values of the other elements each time the selected
element has the desired value, an intersection surface is obtained.
This technique has been used by several authors (e.g. \cite{kat11})
to analyze the structure of phase space in the neighborhood of
stable periodic orbits in a 3D potential, and the properties of the
invariant tori in the 4D spaces of section, under different galactic
potentials. We use it here to study the 3-D NS trajectories and
their 2-D projections.

{We plot the Poincar\'{e} section for $\rm x > 0$, and we fix $\rm y
= 0$ to investigate the dynamical 3-D orbits of NSs, as illustrated
in row C in Fig.~\ref{fig:orbit1}, with varying the initial
parameters. The phase space of NS's motion is 6-D, but since the
total energy and angular momentum are conserved it is in fact only
4-D and its Poincar\'{e} section is 3-D. The initial condition $(x,
y, z, v_{x}, v_{y}, v_{z})$ is reported under the corresponding
column in the Figure. The NSs' motions are very diversified.

\cite{wei10b} investigated the gravitational potential of the
Galactic disk and orbits of stars, and found that all of the orbits
are symmetric with respect to the galactic plane. Here we use the
P90 gravitational potential and find that there are some
non-symmetric orbits, see row D, columns $\beta$ and $\gamma$ in
Fig.~\ref{fig:orbit1}. In row A we can see that when the motion
range in the vertical direction becomes larger than the one in the
radial direction, the orbits become more irregular. See also same
behavior for the projection on x-y plane in row B. While from row E,
we see that the intersection points distribute in some regular lines
on the projection of the Poincar\'{e} section, which is essentially
a closed curve. As such, the motion appears as a quasiperiodic
orbit. However if the motion were exactly periodic, we would expect
that after some time, the star should return back to the same
intersection point on the surface section, and this is not always
the case in our simulations.

\begin{figure*} 
 \begin{center}
 \includegraphics[width=13cm]{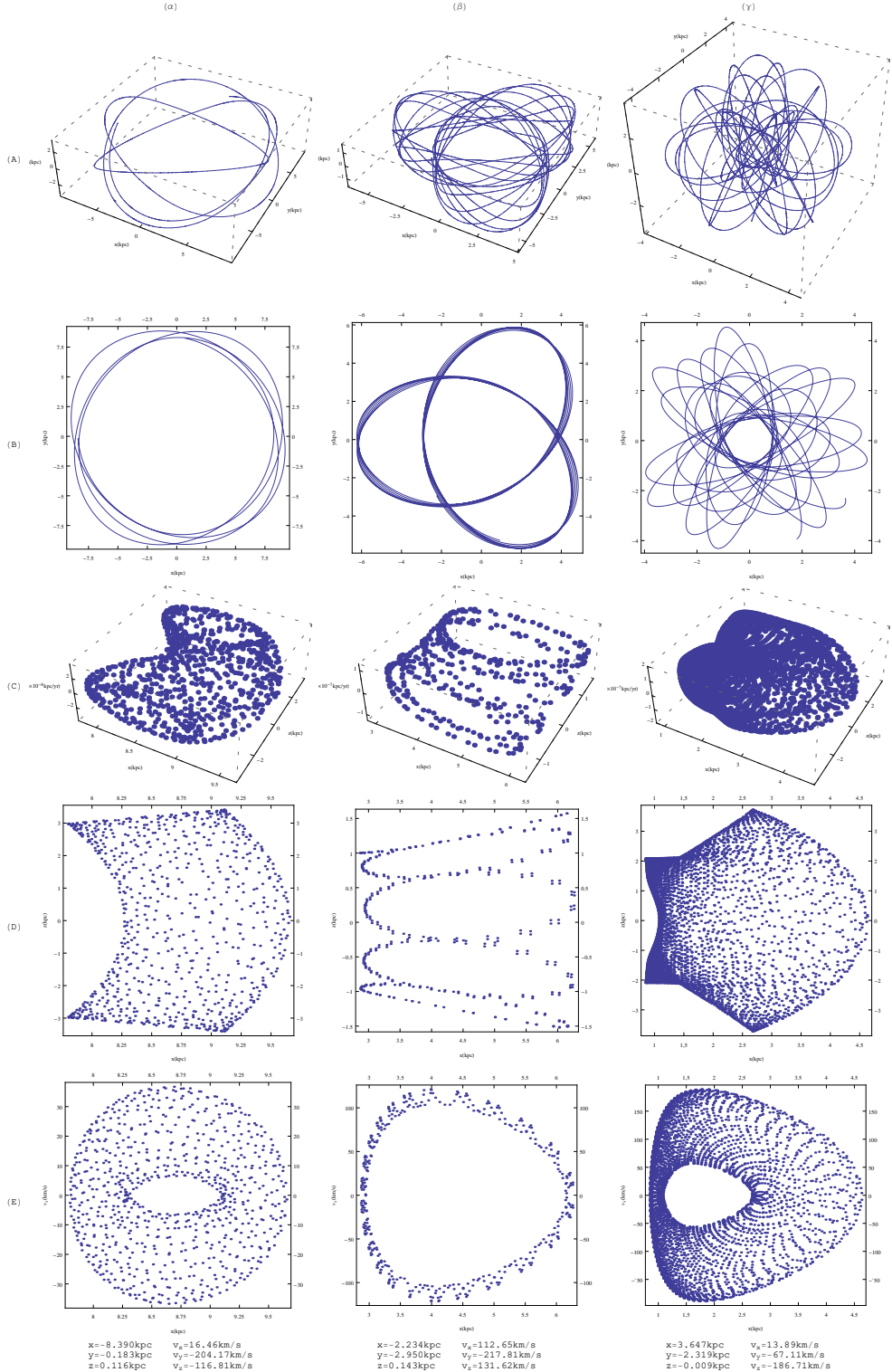}
 \caption{3-D orbits of NSs. Panels in row A: trajectories; panels in row B:
projections of the trajectories on the galactic plane; panels in row C:
Poincar\'{e} section $\rm x>0$, $\rm y=0$ of the 3-D trajectory; panels in row
D: projections on $\rm x$-$\rm z$ plane of the Poincar\'{e} sections; panels in
row E: projections of the Poincar\'{e} sections on $\rm x$-$\rm v_{x}$ plane.}
 \label{fig:orbit1}
 \end{center}
\end{figure*}

According to P90, the dynamical behavior of NS populations is
insensitive to the initial scale-height of progenitors. Here we see
that NS orbits are like those of their progenitors: they are all
basically rotating around the Galactic center, at different radial
distance and uniformly.

\section{Simulation Results and Discussions} \label{sec:rad}

In our calculation we obtain that} the NS distribution is steady
after $10^{9}$~yr. After this time we see that the NSs have greatly
expanded in the radial direction, with the majority of them being
located beyond $25$~kpc from the Galactic center. More precisely
$80\%$ of the old NSs remains within $25$ kpc from the Galactic
rotation axis (i.e. $R<25$ kpc), and $18\%$ instead remains within
$25$ kpc from the Galactic center (i.e. $r<25$ kpc). NSs moving in
and out of the above range are in a dynamically equilibrium state.

\subsection{Radial distribution of NSs}
\begin{figure} 
 \begin{center}
 \includegraphics[width=0.45\textwidth]{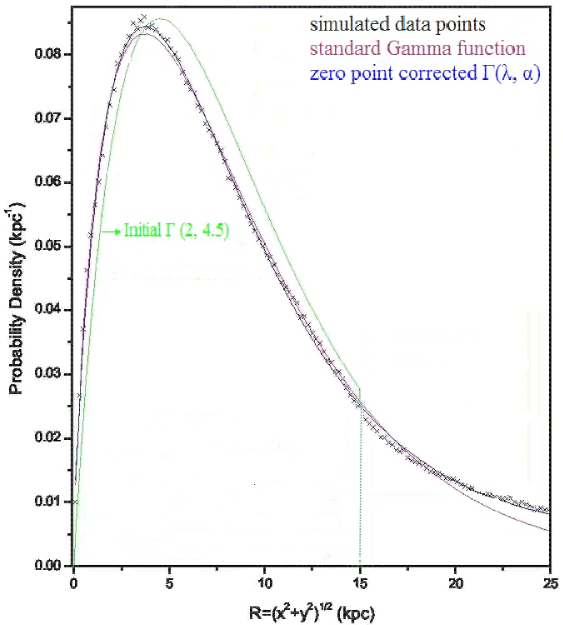}
 \caption{Radial probability density distribution. The crosses (black) indicate
simulated data points; the dotted (green) line indicates the initial
Gamma function; the two solid lines indicate the best fit with the
``zero point corrected $\Gamma(\alpha,\lambda)$" (blue), and with
the standard Gamma function (brown). Details about the fits are
given in Tables~\ref{table:fit2exp}, \ref{table:fitgamma} and
\ref{table:fit0gamma}, respectively.}
 \label{fig:gamma}
 \end{center}
\end{figure}

As we mentioned earlier, NSs are born in the region $0-15$~kpc and
later on they spread to all radii. We follow an approach similar to
\cite{wei10b} in order to investigate the characteristics of old NS
distribution under the two-component Maxwellian initial random
velocity. The normalized position probability density function that
we find is shown in Fig.~\ref{fig:gamma}. We find that the
distribution deviates from the initial distribution, i.e.
$\Gamma(2,4.5)$, due to the NS motion in the Galactic gravitational
field. The distribution peak is now closer to the Galactic center.
At first we fit the normalized $R$ probability density function with
the Gamma function $\Gamma(\alpha,\lambda)$ as:
\begin{equation}
 \label{eq:(8)}
  \rm p_{R}(R)=A\frac{R^{\alpha-1}}{\lambda^{\alpha}}\exp^{-R/\lambda}\;.
\end{equation}
 The best fitting Gamma function is $\Gamma(1.7, 5.2)$. The peak
location of a generic Gamma function $\Gamma(\alpha,\lambda)$ is at
$r_{\rm p} = \lambda(\alpha -1)$. Using this expression for the
initial radial distribution we get $4.5$~kpc, while for the
simulated distribution the peak is at $3.71 \pm 0.17$~kpc. The
fitting results are listed in Tables~\ref{table:fit2exp},
\ref{table:fitgamma} and \ref{table:fit0gamma}.

Due to the unsatisfactory  fitting of Gamma function, specially at
the peak, we use the ``zero point corrected $\Gamma(\alpha,
\lambda)$" proposed by \cite{wei10b}. \textbf{They} modify the Gamma
function $\Gamma(\alpha,\lambda)$ by adding a constant $\rm A_0$:
\begin{equation}
 \label{eq:(9)}
  \rm p_{R}(R) = A_0 + A\frac{R^{\alpha-1}}{\lambda^{\alpha}}\exp^{-R/\lambda},
\end{equation}

If the points are close to a Gamma distribution function then the
scatter will be small relative to the total variation in the values
of the response variable. We adopt the coefficient of determination
(COD, also known as r-squared) to measure the fit quality. The
closer COD to 1, the better the fit. Figure~\ref{fig:gamma}
indicates that the Gamma distribution function is quite satisfactory
with a $\rm COD$ of 0.99, and Eq.~(\ref{eq:(9)}) is acceptable for
the case with the relative standard errors of the fitting parameters
less than $5\%$.

As for the $\Gamma$ function, also for the ``zero point corrected
$\Gamma$'' the peak position is at $\lambda(\alpha -1 )$ and for the
best fit case it is at $3.73 \pm 0.19$~kpc. We notice that the
evolution of the NSs in the Galactic gravitational field makes the
value of their $R$-distribution at $R = 0$~kpc not exactly equal to
0. In other words, there is a ``zero shift", which is the total
effect of the NS orbits shown in Sec.~\ref{sec:orb}

\subsection{Vertical distribution of NSs}
We consider the vertical distribution of the bound NSs in the whole
Galactic disk with $R < 25$~kpc and we find that it is not well
described by a single exponential decay. For this reason we employ a
double exponential profile:
\begin{align}
 \label{eq:(11)}
  \rm p_{z}(z) = \emph{A}_{0} \times g(z) &+ A_{1}\times \exp\left[\rm -z/h_{1}\right] \nonumber \\
                &+ A_{2}\times \exp\left[\rm -z/h_{2}\right],
\end{align}
where $\rm A_{0}\times$ g(z) represents the disk component, g(z) is
step function which is 1 in the disk and 0 outside, and h$\rm _1$
and h$\rm_2$ are the height scales of the two exponential
contributions. Without loss of generality we can assume h$\rm_{1} <
h_{2}$ and refer to A$_{1}\rm \times \exp\left[-z/h_{1}\right]$ as
the low-scale-height component and to A$_{2}\rm\times
\exp\left[-z/h_{2}\right]$ as the high-scale-height component. The
probabilities for the low-scale-height and the high-scale-height
component are respectively:
\begin{align}
\rm P_{1}= \int_{0}^{\infty} A_{1}\times \exp\left[-z/h_{1}\right] dz = A_{1}
\times h_{1} \\
\rm P_{2}= \int_{0}^{\infty} A_{2}\times \exp\left[-z/h_{2}\right]
dz = A_{2} \times h_{2} .
\end{align}

We also study the half density scale height of the disk
z$\rm_{1/2}$, defined as the height at which the \textbf{total}
probability density drops to $50\%$ of the Galactic plane one. These
results are shown in Table~\ref{table:fit2exp}. We get a COD
$\sim0.99$ and relative standard errors $\lesssim 1\%$, except that
the relative standard error of $\rm p_{0}$ lies in the range
$6.4~\%$.

\begin{table}[ht]
 \caption[]{Parameters of two exponential decay model Eq.~(\ref{eq:(11)}).}
 \label{table:fit2exp}
\begin{center}
\begin{tabular}{lcc}
\hline \hline \noalign{\smallskip}
parameter & value &  relative error $\%$ \\
\hline \noalign{\smallskip}
A$_{0}\rm (kpc^{-1})$    & $1.8\times 10^{-5}$  &6.4\\
\noalign{\smallskip}
A$_{1}\rm (kpc^{-1})$   & 1.87 &  0.02   \\
\noalign{\smallskip}
h$_{1}\rm (kpc)$     & $20.6\times 10^{-3}$   &0.04 \\
\noalign{\smallskip}
A$_{2}\rm (kpc^{-1})$ & $35.6\times 10^{-3}$ & 0.09   \\
\noalign{\smallskip}
h$_{2}\rm (kpc)$   & 1.55 & 0.08    \\
\noalign{\smallskip}
COD   & 0.999 & -    \\
\noalign{\smallskip}
P$\rm _{2}/P_{1}$       & 1.45 &  0.003  \\
\noalign{\smallskip}
z$\rm_{1/2} (kpc)$   & $17.6\times 10^{-3}$ & 0.01  \\
\noalign{\smallskip} \hline
\end{tabular}
\end{center}
\end{table}

\begin{table}[ht]
 \caption[]{Best fit results for the standard Gamma function.}
 \label{table:fitgamma}
\begin{center}
\begin{tabular}{lcc}
\hline \hline \noalign{\smallskip}
parameter & value & relative error $\%$ \\
\hline \noalign{\smallskip}
\noalign{\smallskip}
$\rm A$   & 1.13 &  0.49   \\
\noalign{\smallskip}
$\rm \alpha$     & 1.71 &  $0.54$   \\
\noalign{\smallskip}
$\rm \lambda (kpc)$ & 5.21 & 0.86   \\
\noalign{\smallskip}
COD  & $0.996$ & -  \\
\noalign{\smallskip} \noalign{\smallskip} \hline
\end{tabular}
\end{center}
\end{table}

\begin{table}[ht]
 \caption[]{Best fit parameters for the ``zero point corrected
$\Gamma(\alpha,\lambda)$".}
 \label{table:fit0gamma}
\begin{center}
\begin{tabular}{lcc}
\hline \hline \noalign{\smallskip}
parameter & value & error $\%$ \\
\hline \noalign{\smallskip}
A$\rm_{0} (kpc^{-1})$    & $5\times10^{-3}$ & 6.3  \\
\noalign{\smallskip}
$\rm A$   & $95.6\times10^{-3}$ &  1.2   \\
\noalign{\smallskip}
$\rm\alpha$     & 1.83 &  0.58   \\
\noalign{\smallskip}
$\rm \lambda (kpc)$ & 4.48 & 1.1   \\
\noalign{\smallskip}
COD  & 0.998 & - \\
\noalign{\smallskip} \hline
\end{tabular}
\end{center}
\end{table}

\begin{table}[ht]
 \caption[]{Best fit parameters for linear relations of the radial distributions
of the two scale heights $\rm h_1$ and $\rm h_2$.$^{\dag}$ is for
fitting the region inward of $4.25$~kpc and $^{\ddag}$ is for
fitting the region outward.}
 \label{table:fith}
\begin{center}
\begin{tabular}{lcc}
\hline \hline \noalign{\smallskip}
parameter & value & error $\%$ \\
\hline \noalign{\smallskip}
k$\rm_{1}$    & $13\times10^{-3}$ & $0.26$  \\
\noalign{\smallskip}
b$\rm_{1}$   & $12.8\times10^{-3}$  &   $0.005$ \\
\noalign{\smallskip}
k$\rm_{2}^{\dag}$     & $18.4\times10^{-3}$  &  0.0007      \\
\noalign{\smallskip}
b$\rm_{2}^{\dag}$ & 0.03  & 0.002   \\
\noalign{\smallskip}
k$\rm_{2}^{\ddag}$ & 0.05   & 0.66  \\
\noalign{\smallskip}
b$\rm_{2}^{\ddag}$ &  0.65  &  0.16 \\
 \noalign{\smallskip} \hline
\end{tabular}
\end{center}
\end{table}

Observational studies of the Galactic disk reveal that it can be
well described in terms of the two components model with a thin disk
and a thick disk component (see e.g. \cite{che01}, \cite{kae05},
\cite{wei10a}). Our simulation confirms the validity of this model
and shows that the $\rm z$ hierarchy effect can be regarded as the
result of the dynamical evolution of the old NSs originated from the
Galactic disk.

\subsection{Scale-Height vs $R$ relation}
We now consider the $\rm z$ distribution at different Galactic
radial distances $\rm R$ from the Galactic center. To this end, we
divide $\rm R$ with $0.5$ ~kpc spacing from $0$ ~kpc to $25$ ~kpc
and get $50$ parts, then we analyze in details the NS $\rm z$
distribution in each part. The two exponential decay of
Eq.~(\ref{eq:(11)}) is  still employed to study the case in each
spacing segment of $\rm R$, to derive the high-scale-heights $\rm
h_{2}$, low-scale-heights $\rm h_{1}$,  and the ratios of the two
components $\rm P_{2}/P_{1}$.

\begin{figure} 
 \begin{center}
 \includegraphics[width=0.45\textwidth]{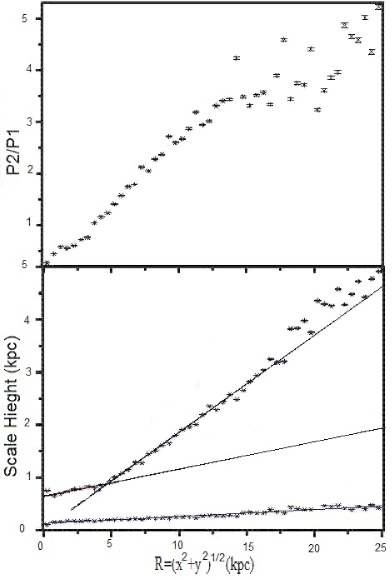}
 \caption{Scale-heights in the double exponential decay model
see Eq.~(\ref{eq:(11)}). Top panel: ratio of $\rm P_{2}/P_{1}$.
Bottom panel: radial distributions of the two scale heights $\rm
h_1$ and $\rm h_2$. For both of the panels the crosses indicate
simulated data points, while the solid lines indicate the best fit
line. The fitting parameters are given in Table~\ref{table:fith}.}
 \label{fig:scaleh}
 \end{center}
\end{figure}

The results are shown in Fig.~\ref{fig:scaleh}. In each segment of
$\rm R$ the two exponential model is still significantly effective.
As a result, the $\rm COD$s of both distributions in
Fig.~\ref{fig:scaleh} is quite similar $\sim$ 0.99, and  with very
small relative standard errors $\lesssim 0.01$ . The fitting results
are listed in Table~\ref{table:fith}.

\begin{figure} 
 \begin{center}
 \includegraphics[width=0.45\textwidth]{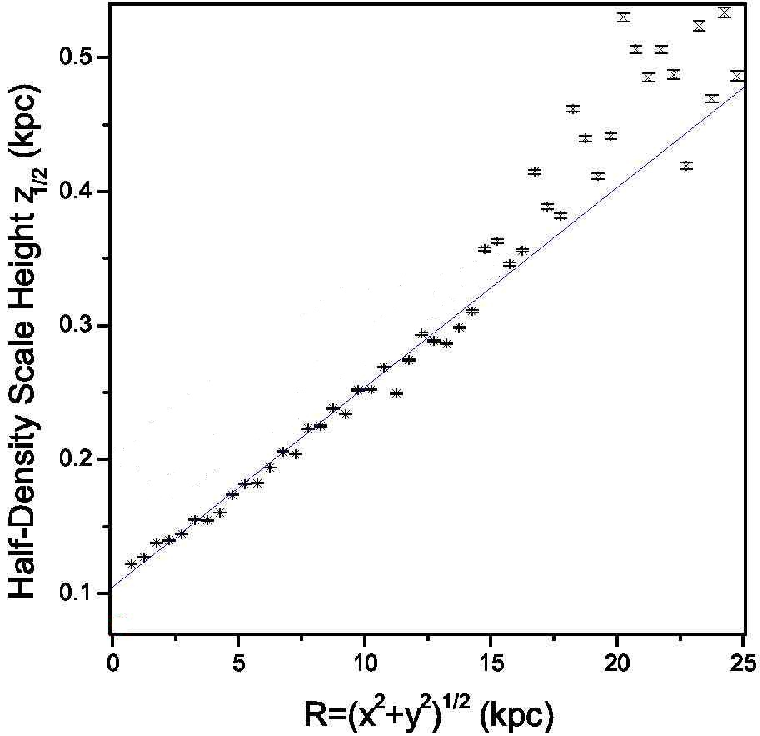}
 \caption{Radial distributions of the half density scale heights in
the region R$\rm <25$~kpc with the fitting parameters of
Eq.~(\ref{eq:(12)}) for the straight lines.}
 \label{fig:h12}
 \end{center}
\end{figure}

The relationship between the two scale heights and $\rm R$ can be described by
the following linear model.
\begin{equation}\label{eq:(12)}
  \rm h(R) = k\times R+b,
\end{equation}
 We also plot the radial dependence of the half
density scale heights in Fig.~\ref{fig:scaleh}. The fitting results
of parameters $\rm k$ and $\rm b$ are listed in Table~\ref{fig:h12}.

The low-scale-heights can be depicted by a linear function of $\rm
R$ within the whole range of the Galactic disk $0-25$ ~kpc. The
slope of the fitting line is small, which means that the changing of
this component within $0-25$ ~kpc of $\rm R$ is not large.

For the high-scale-height components, there exists a point at $R =
4.25$~kpc, where the behavior changes. Both sides of this point have
a linear radial dependence but with different slopes: the one inside
$R < 4.25$~kpc is smaller than the one outside $R > 4.25$~kpc. We
notice that $R = 4.25$~kpc corresponds to the observed $R_0$ of the
HI disk.

As regards the $\rm R$ distribution of $\rm P_{2}/P_{1}$, we can see
the difference of old NS distribution under \cite{ARal2002} from
those under \cite{hob05} and \cite{fau06} in \cite{wei10b}. Where
the $\rm R$ distributions of $\rm P_{2}/P_{1}$ have three distinct
parts clearly, and the high-scale-height distributions have not
points like the observed R$\rm_{0}$ of the HI disk. The ratio of the
two components
 leads to the increase of growth slowly and smoothly with $\rm R$ in
the whole Galaxy.

\begin{table}[ht]
 \caption[]{Best fit parameters for linear relations of the radial distributions
of the half density of two scale heights Eq.~~(\ref{eq:(12)}).}
 \label{table:fith12}
\begin{center}
\begin{tabular}{lcc}
\hline \hline \noalign{\smallskip}
parameter & value & error $\%$ \\
\hline \noalign{\smallskip}
k    & $14.9\times10^{-3}$ & $0.01$  \\
\noalign{\smallskip}
b   & $10.5\times10^{-3}$ &  0.01   \\
\noalign{\smallskip} \hline
\end{tabular}
\end{center}
\end{table}

\clearpage

The NS scale-heights generally increase from the Galactic center to
the edge of Galactic disk. This phenomenon is independent of the
initial velocity distribution of the NSs, and it is due to the
action of the Galactic gravitational field on the NSs. Our
calculation shows that the heights of the orbits of the NSs
generally decrease towards the Galactic center (see
Sec.~\ref{sec:orb}). P90 calculated the half density scale height in
the vicinity of the Sun (R$\rm_{0}=8$~kpc) and obtained z$\rm
_{1/2}=0.2$~kpc. In our present updated version the corresponding
value is $0.2\pm0.0004$~kpc.

\section{Discussions and Conclusions} \label{sec:rest}
In this paper we have investigated the space and velocity
distribution of old neutron stars (NSs) in our Galaxy. We assume
that the initial velocity distribution is the result of three
components: the kick velocity, the circular velocity and the
Maxwellian velocity (following \cite{ARal2002}). For the initial position
distribution instead we assume that it follows a $\Gamma$ function.
As regards the Galactic gravitational potential, we follow the \cite{pac90}
prescription, which is of course only and approximation, since it
does not take into account any inhomogeneity within the Galaxy.
However this is suitable for the simplified analysis that we are
developing here. We have then used Monte Carlo simulations to let
the NSs evolve and have shown 3-D NS orbits and Poincar\'{e}
sections of the phase space.

It is evident that the irregular character of the motion of NSs
increases when the vertical direction becomes larger than radial
direction. Another remarkable finding is that there are some
significant diffractions in the symmetric of the orbits, which may
effects of supernovae kicks.

%

Our numerical results show that NSs have a very large radial
Galactic expansion. The majority of them ($80\%$) falls within $25$
kpc from the Galactic rotation axis ($R<25$ kpc), and $18\%$ instead
remains within $25$ kpc from the Galactic center ($r<25$ kpc). An
important aspect is that the total number of NSs moving in and out
of the above range is in a dynamically equilibrium state after
$10^9$ yr.

The radial probability density distribution deviates from the
initial distribution, and has a peak which is closer to the Galactic
center. The analysis of the vertical and radial distributions
clearly show that  the orbits of the NSs decrease toward the
Galactic center within different scale heights.

Qualitatively, the simulated old NSs disk, especially the middle and
outer components, keeps the observed HI disk in moderation. Although
the old NSs and their progenitors have different radial and vertical
distribution, we find that the shapes of their orbits are quite
similar in the HI clouds regions.

The results of this work will constitute the base for further
studies on NS properties. Such research could be helpful for the
detection of old NSs via their gravitational microlensing that
result in the variation in the brightness of the distant active
galactic nuclei (e.g. ~\cite{haw93, haw02}). Another way for
detecting the old NSs is through the interaction with the
interstellar medium \citep{Pop02}.

As subsequent steps we plan to (1) apply the three exponential decay model in
studying the NS vertical distribution with more detail and (2) to use different
models of the Galactic potential to investigate specific parts of our Galaxy.

\section*{Acknowledgments}

This work is supported by the National Natural Science Foundation of
China (NSFC 10773017, NSFC 10773034) and National Basic Research
Program of China (2009CB824800, 2012CB821800). Chinese Academy of Sciences and
National Astronomical Observatory of China (NAOC) of CAS has supported this work
by the Silk Road Project (CAS Grant Number 2009S1-5). L.N. is currently
supported by a Chinese Academy of Sciences fellowship for young international
scientists (Grant Number 2010Y2JB12).


\begin{thebibliography}{99}
%
\bibitem[Arzoumanian et al.(2002)]{ARal2002}
Arzoumanian, Z., Chernoff, D.~F., \& Cordes, J.~M.\   Astrophys. J.
\textbf{568}, 289 (2002)
%
\bibitem[Bhattacharya  \& van den Heuvel (1991)]{BH91}
Bhattacharya D. \& van den Heuvel E. P. J., Phys. Rep., \textbf{203}, 1 (1991)
%
\bibitem[Binney  \& Merrifield (1998)]{bin98}
Binney J. \& Merrifield M., Galactic Astronomy (Princeton:
Princeton University Press) (1998)
%
\bibitem[Caldwell \& Ostriker (1981)]{cal81}
Caldwell J. \& Ostriker J., Astrophys. J. \textbf{251}, 61 (1981)
%
\bibitem[Chen et al. (2001)]{che01}
Chen B., Stoughton C., Smith J. A., et al. Astrophys. J. \textbf{553}, 184
(2001)
%
\bibitem[Cordes \& Chernoff (1998)]{cor98}
Cordes J. M., \& Chernoff D. F., Astrophys. J. \textbf{505}, 315 (1998)
%
\bibitem[Cordes \& Lazio(2002)]{CL2002}
Cordes, J.~M., \& Lazio, T.~J.~W.\  arXiv:astro-ph/0207156 (2002)
%
\bibitem[Faucher-Gigu\`{e}re \& Kaspi (2006)]{fau06}
Faucher-Gigu\`{e}re C.-A. \& Kaspi V. M., Astrophys. J. \textbf{643}, 332
(2006)
%
\bibitem[Frei et al.(1992)]{fre92}
Frei Z., Huang X. \& Paczy\'{n}ski B., ApJ. \textbf{643}, 332 (1992)
%
\bibitem[Gonthier et al.(2002)]{gon02}
Gonthier P. L., Ouellette M. S., Berrier J. et al. Astrophys. J. \textbf{565},
482 (2002)
%
\bibitem[Gott et al.(1970)]{gott70}
Gott J. R., Gunn J. E., \& Ostriker J. P., Astrophys. J. \textbf{160}, L91
(1970)
%
\bibitem[Hansen \& Phinney (1997)]{han97}
 Hansen B. M. S. \& Phinney E. S., Mon. Not. R. Astron. Soc. \textbf{291},
569 (1997)
%
\bibitem[Hawkins (1993)]{haw93}
Hawkins M.R.S., Nature \textbf{366}, 242 (1993)
%
\bibitem[Hawkins (2002)]{haw02}
Hawkins M.R.S.,  Mon. Not. R. Astron. Soc. \textbf{329}, 76 (2002)
%
\bibitem[Hobbs et al.(2005)]{hob05}
Hobbs G., Lorimer D. R., Lyne A. G., \& Kramer M., Mon. Not. R.Astron. Soc.
\textbf{360}, 974 (2005)
%
\bibitem[Jones \& Lambourne (2004)]{Jon04}
Jones M. H, \& Lambourne R.J.A. An Introduction to Galaxies and
Cosmology. Cambridge: Cambridge University Press, 7 (2004)
%
\bibitem[Kaempf et al. (2005)]{kae05}
Kaempf T. A., de Boer K. S., \& Altmann M., Astron. \& Astrophys,
\textbf{432}, 879 (2005)
%
\bibitem[Katsanikas \& Patsis (2011)]{kat11}
Katsanikas M. \& Patsis P. A., Int. J.Bif. Chaos, \textbf{21}, 467 (2011)
%
\bibitem[Lorimer (2008)]{lor08}
Lorimer D. R.  Living Rev. Relativity, \textbf{11}, 8 (2008)
http://relativity.livingreviews.org/Articles/lrr-2008-8/
%
\bibitem[Lyne et al. (1982)]{lyn82}
Lyne A. G., Anderson B., \& Salter M. J., Mon. Not. R. Astron. Soc.
\textbf{201}, 503 (1982)
%
\bibitem[Lyne et al.(2004)]{lyn04}
Lyne A. G., Burgay M., Kramer M. et al.  Science, \textbf{303}, 1153 (2004)
%
\bibitem[Lyne \& Graham-Smith (2007)]{lyn07}
Lyne A. G. \& Graham-Smith, F., Pulsar Astronomy, Cambridge Astrophysics Series,
Cambridge University Press, 3ed Edit. (2007)
%
\bibitem[Mdzinarishvili \& Melikidze (2004)]{Mdz04}
Mdzinarishvili T. G. \& Melikidze G. I., Astron.\& Astrophys.,
\textbf{425}, 1009 (2004)
%
\bibitem[Ofek (2009)]{Ofe09}
Ofek E. O., PASP, \textbf{121}, 814 (2009)
%
\bibitem[Paczy\'{n}ski (1990)]{pac90}
Paczy\'{n}ski B., Astrophys. J. \textbf{348}, 485 (1990)
%
\bibitem[Patsis \& Grosb{\o}l (1996)]{Pat96}
Patsis P. A. \& Grosb{\o}l P., Astron.\& Astrophys., \textbf{315},
371 (1996)
%
\bibitem[Patsis et al.(2002)]{Pat02}
Patsis P. A., Athanassoula E., Grosb{\o}l P. et al. Mon. Not. R.
Astron. Soc. \textbf{355}, 1049 (2002)
%
\bibitem[Perets et al.(2009)]{Per09}
Perets H. B, Wu X., Zhao H. S. et al. ApJ, \textbf{697}, 2097 (2009)
%
\bibitem[Popov et al.(2000)]{Pop02}
Popov S. B., Colpi M., Treves A. et al. ApJ, \textbf{530}, 896
(2000)
%
\bibitem[Sartore et al.(2010)]{sar10}
Sartore N., Ripamonti E., Treves A. \& Turolla R.,  Astron.\& Astrophys.,
\textbf{510}, A23 (2010)
%
\bibitem[Taylor \& Cordes  (1993)]{Tay93}
Taylor J. H., \& Cordes  J. M., Astrophys. J. \textbf{411}, 674
(1993)
%
\bibitem[Wei et al.(2010a)]{wei10a}
Wei Y. C., Taani A., Pan Y. Y. et al.,  Chin. Phys. Lett., \textbf{27}, 9801
(2010a)
%
\bibitem[Wei et al.(2010b)]{wei10b}
Wei Y. C., Chengmin C. M., Xinji W. et al.,  Scince in China, 53,
1939 (2010b)
%
\end{thebibliography}


\end{document}